\documentclass[twocolumn,5p]{elsarticle}

\usepackage{amsmath}
\usepackage{amssymb}
\usepackage{graphicx}
\usepackage{hyperref}
\usepackage{lineno}

\biboptions{sort&compress}

\begin{document}

\begin{frontmatter}

\title{Chiral states around a mass-inverted quantum dot in graphene}
\author{Nojoon Myoung}
\address{Department of Physics Education, Chosun University, Gwangju 61452, Republic of Korea}
\ead{nmyoung@chosun.ac.kr}

\begin{abstract}
Topologically protected chiral states at a mass-inverted quantum dot in graphene are studied by analyzing both tight-binding and kernal polynomial method calculations. The mass-inverted quantum dot is introduced by considering a heterojunction between two different mass domains, which is similar to the domain wall in bilayer graphene. The numerical results show emergent metallic channels across the mass gap when the signs of the mass terms are opposite. The eigenstates of the metallic channels are revealed to be doubly degenerate---each state propagates along opposite directions, maintaining the time-reversal symmetry of graphene. The robustness of the metallic channels is further examined, concluding with the fact that chiral states are secured unless atomic vacancies form near the domain wall. Such chiral states circulating along the topological defects may pave a novel route to engineering topological states based on graphene.
\end{abstract}


\end{frontmatter}

\section{Introduction}

In the past decade, extensive research has pursued topological edge states in electronic \cite{Kane2005,Bernevig2006,Martin2008,Jung2011,Yang2012,Chang2013,Vaezi2013,Zhang2013,Drozdov2014,Li2014,Qiao2014,Du2015,Ju2015,Li2016,Zhu2019,Hu2019,Gou2020}, photonic \cite{Hafezi2013,Barik2016,StJean2017,Wu2017,Xiao2017,Noh2018,Han2019,Olekhno2020,Yu2020}, acoustic \cite{Ni2019,Gao2020,Wang2020,Xia2020}, and atomic systems \cite{Goldman2013}, as well as lattice models \cite{Bauer2018,Liu2019a,Obana2019,Smirnova2019,Zhang2019,Ezawa2020,Song2020}. As candidate materials for research on topological natures, graphene and bilayer graphene (BLG) possess a crucial advantage: the tunability of topological states via electrostatic gating. Actually, since the first proposal of one-dimensional (1D) topological states at a domain wall in oppositely biased BLG \cite{Martin2008}, there has been considerable progress in understanding both physical and topological characteristics of the topological states \cite{Vaezi2013,Ju2015,Yin2016,Liu2019b,Wang2019,Ma2020,Jiang2020,Chou2020}. For example, it has been theoretically proven that the formation of topological interface states at the domain wall in BLG follows from a Chern number inversion according to gate-polarity \cite{Vaezi2013}. In fact, the same topological phenomenon has also been demonstrated in gapped single-layer graphene by considering mass-inverted domain walls \cite{Lado2013,Qiao2014}. However, so far, relatively less attention has been paid to the topological states at mass-inverted interfaces in single-layer graphene compared to those in BLG.

The heterojunction in gapped graphene with inhomogeneous mass terms provides marvelous approaches to hosting topologically protected metallic channels. Such an inhomogeneous mass gap distribution in graphene is inevitable due to periodic interlayer coupling with substrates \cite{Ratnikov2009,Rusponi2011,Kim2018}. Here, we report the existence of chiral modes around a mass-inverted quantum dot (MIQD) in gapped graphene. The chiral states are doubly degenerate, with each state circulating along the boundary of the MIQD either in counterclockwise or clockwise direction. We also show a metal-to-insulator transition as the mass gap distribution changes, corresponding to the gap closing with the emergent topological channels. Further, we demonstrate that the doubly degenerate chiral states are robust against some perturbations, such as continuous mass inversion or atomic vacancies, which may occur in realistic situations.

\section{Model: Mass-inverted quantum dot}

\begin{figure}
\centering
\includegraphics[width=5.5cm]{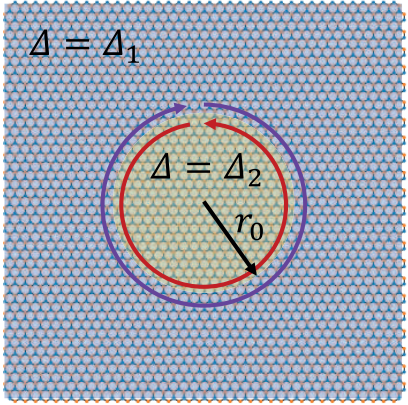}
\caption{Schematic model of the system considered in this study. Different mass gaps $\Delta_{1,2}$ are given outside and inside the MIQD (yellow region) with radius $r_{0}$, respectively. Counterclockwise and clockwise chiral modes emerge along the MIQD boundary only when the signs of the two mass gaps are opposite.}
\label{fig:model}
\end{figure}

To describe the MIQD, we start with a continuum Hamiltonian for the K valley,
\begin{align}
H=v_{F}\vec{\sigma}\cdot\vec{p}+\sigma_{z}\Delta\left(r\right),
\end{align}
where $v_{F}\simeq 10^{6}$ ms$^{-1}$ is the Fermi velocity of graphene, $\vec{\sigma}$ are Pauli matrices, and
\begin{align}
\Delta\left(r\right)=\left\{\begin{array}{ll}\Delta_{1},&r>r_{0}\\\Delta_{2},&r<r_{0}\end{array}\right.,
\end{align}
where $r_{0}$ is the radius of the MIQD as depicted in Fig. 1. In the Hamiltonian, $\sigma_{z}$ acting on the sublattices implies that $\Delta_{1,2}$ are staggered potentials breaking sublattice symmetry, so that a finite mass gap opens near the charge neutrality point. This resulting mass gap will be twice $\Delta_{1,2}$ in each domain. The lack of both intervalley and spin-orbit interactions leads the electronic states of the system to be valley- and spin-independent, and thus the continuum model is reduced into the simplest case, i.e., spin and valley degrees of freedom are degenerate. While the energy spectra above and below the gap (conduction and valence bands) can be studied by solving the analytic solutions of the continuum Hamiltonian via wavefunction matching conditions \cite{VanPottelberge2017,Belouad2018,Xu2020}, numerical studies are required to explore the eigenenergies and eigenstates inside the mass gap. Thus, in this study, numerical calculations based on the tight-binding Hamiltonian including discretized S-matrix and the kernel polynomial method (KPM) are performed using \textsc{kwant} codes \cite{Groth2014}.

Before presenting the results, it is apposite to briefly recall the topological origin of the interface states at the domain wall with opposite mass gaps. While gapped graphene with sublattice-symmetry-breaking is topologically nontrivial, its Chern number is zero due to the even number of Dirac cones in the Brillouin zone. In fact, the Chern number of gapped graphene around each valley (the so-called valley Chern number) is given by $\mathcal{C}_{K}=\mbox{sgn}\left(\Delta\right)/2$ and $\mathcal{C}_{K'}=-\mbox{sgn}\left(\Delta\right)/2$, resulting in a total Chern number of zero, as $\mathcal{C}=\mathcal{C}_{K}+\mathcal{C}_{K'}=0$ \cite{Wang2014}. Such a topological nature of gapped graphene is in good agreement with the lack of topological edge states. However, if there exists a heterojunction formed by two mass-inverted regions, one can expect two topological edge states along the junction that propagate in opposite directions to each other, like the domain walls formed in opposite-gate BLG and twisted BLG \cite{Martin2008,Ju2015,Yin2016}.

\section{Results and Discussion}

\subsection{Chiral edge states of the MIQD}

\begin{figure*}
\centering
\includegraphics[width=15cm]{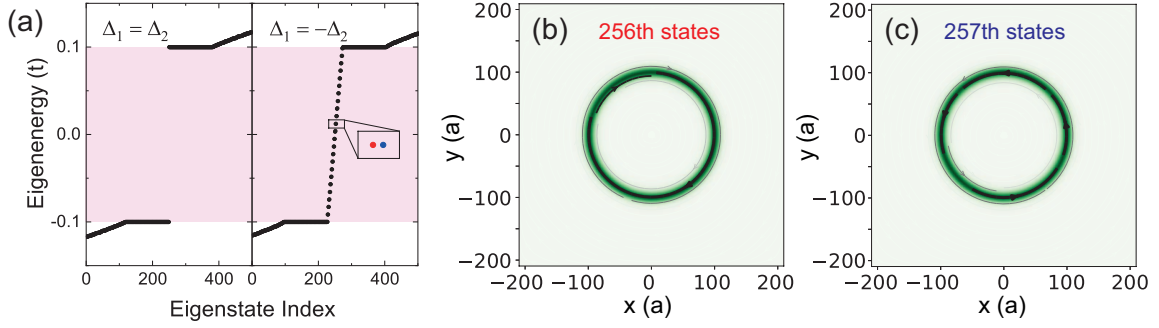}
\caption{(a) Eigenenergy spectra of the system for a homojunction (left panel) and heterojunction (right panel) of the mass terms. The mass gap of the system is indicated by the shaded area. The inset in the right panel is a zoom-in of a pair of degenerate eigenstates. (b) and (c) Densities and currents of the eigenstates in the inset of (a). Dark and light green represent higher and lower densities, and the solid arrows indicate the current direction.}
\label{fig:evals}
\end{figure*}

First, eigenenergies and eigenstates are calculated to confirm the emergence of the topological edge states around the MIQD. Square-shaped closed systems are considered with a side $L=400~a$, where the MIQD of radius $r_{0}=100~a$ exists ($a$ is the lattice constant of graphene), as depicted in Fig. \ref{fig:model}. For convenience, only 500 eigenstates around zero energy are calculated. For homogeneously gapped graphene, i.e. $\Delta_{1}=\Delta_{2}$, there is no eigenenergy in the range from $-0.1\Delta_{1}$ to $+0.1\Delta_{1}$ which is a mass gap induced by the staggered potential, as highlighted by the shaded area in Fig. \ref{fig:evals}(a). On the other hand, when $\Delta_{1}=-\Delta_{2}$, a number of eigenstates exist inside the mass gap, of which eigenenergies fill the mass gap out---this means that the system undergoes a transition from an insulator to a metal via the mass inversion.

Meanwhile, it is clearly seen that the mid-gap eigenstates are doubly degenerate, as shown in Fig. \ref{fig:evals}(a). A couple of degenerate eigenstates are selected [see the inset of Fig. \ref{fig:evals}(a)], and their spatial distributions are plotted in Figs. \ref{fig:evals}(b) and (c). As expected, these states are strongly localized along the MIQD edge, and their propagating directions are opposite to each other. Such chiral MIQD states are robust against geometrical deformation of the dot, since they are topologically protected \cite{Xu2020}.

\begin{figure}
\centering
\includegraphics[width=\linewidth]{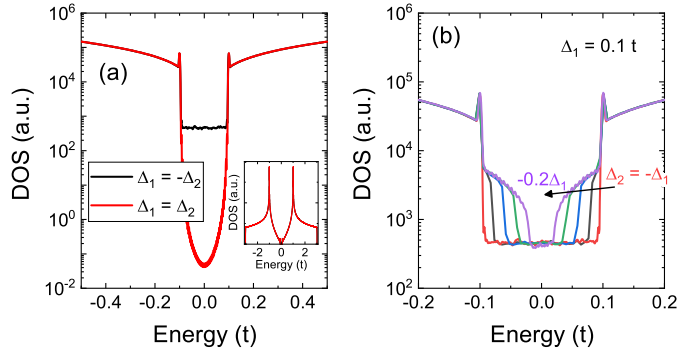}
\caption{Electronic properties of the system calculated using KPM. (a) DOS spectra near the mass gap as functions of energy for the MIQD (black curve) and for the homogenously gapped graphene (red curve). Inset: Full DOS spectra over the entire energy range of the system. (b) DOS spectra near the mass gap as functions of energy for various $\Delta_{2}$ values.}
\label{fig:dos}
\end{figure}

As aforementioned, the emergence of the topological states in the MIQD corresponds to the metal-to-insulator transition via mass term variation. In order to demonstrate the transition behavior, the electronic properties of the system are studied by calculating the density of states (DOS) with the KPM using \textsc{kwant} codes. The closed system for the KPM is a square-like graphene with a side $L=1000~a$ where an MIQD of radius $r_{0}=200~a$ is introduced. Figure \ref{fig:dos}(a) exhibits the distinguishable DOS spectra between two cases: the MIQD (black curve) and homogeneously gapped graphene (red curve). Even in the mass gap, the DOS of the MIQD is approximately $10^{4}$ times greater than that of the homogeneously gapped graphene. The DOS plateau of the MIQD in the mass gap regime is correlated to the fact that 1D channels are employed in the electronic properties.

Besides, the electronic properties of the MIQD are entirely insensitive to the imbalance of the mass terms. As shown in Fig. \ref{fig:dos}(b), the DOS remains unchanged even when the magnitude of $\Delta_{2}$ differs from $\Delta_{1}$, simply narrowing the energy gap of the system as $\left|\Delta_{2}\right|$ decreases. Let us also notice that the DOS spectra are achieved by averaging results from ten individual KPM computations, since the KPM relies on random vector sets to solve linear systems \cite{Weisse2008}.

Next, the robustness of the topological states is discussed for two types of perturbations: atomic vacancies and continuous mass inversion. Both perturbations would likely occur in realistic devices in experimental studies.

\subsection{Robustness of the chiral states I: Vacancies}

\begin{figure}
\centering
\includegraphics[width=8.5cm]{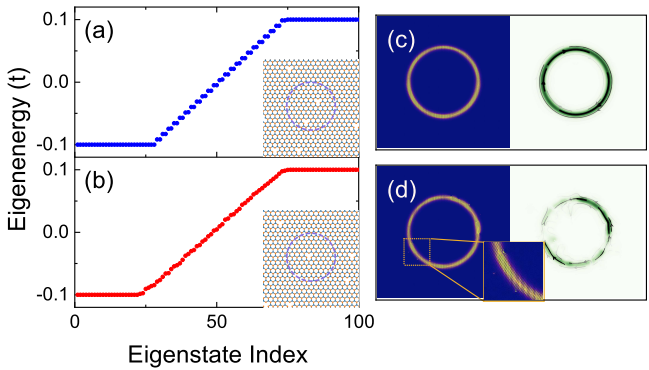}
\caption{Effects of carbon vacancies on the MIQD in graphene. (a) and (b) Eigenenergy spectra of the system with vacancy distribution restricted near the domain wall (a) and distributed over the whole system (b). (c) and (d) Density and current maps of the 48th eigenstate from the spectra in (a) and (b), respectively.}
\label{fig:vacancy}
\end{figure}

First, let us investigate the effects of atomic vacancies on the chiral states when $\Delta_{1}=-\Delta_{2}$. It has been reported that atomic vacancies host localized states, causing valley symmetry breaking\cite{Libisch2011,RuizTijerina2016}. As discussed before, the emergence of the chiral states originates from the discontinuity of the valley Chern number at the domain wall, i.e., $\Delta\mathcal{C}_{K}=+1$ and $\Delta\mathcal{C}_{K'}=-1$, which correspond to the opposite propagating directions. Thus, the vacancy-induced valley symmetry breaking is expected to lift up the degeneracy of the topological channels. Considering the fact that the defect-induced localization is short-range in atomic scale, two separate systems are examined: one in which the vacancy distribution is restricted in the vicinity of the domain wall, and one in which the vacancies are randomly distributed over the whole system. Particularly, the latter case contains an interesting situation that vacancies are formed quite close to the domain wall, whereas the chiral states cannot `see' the vacancies in the former case. Indeed, Fig. \ref{fig:vacancy} shows that the chirality of the topological states can deteriorate only when vacancies exist near the domain wall. It is also observed that the eigenstates are not doubly degenerate any longer, accompanied by a chirality breaking, as shown in Fig. \ref{fig:vacancy}(b). 

This finding is in line with the fact that localization at a vacancy is atomically short-range, so any changes due to the vacancy-hosting localized states are local effects, with no influence over long distance. Although the localized states at a vacancy locally break the valley symmetry, the chiral states are unaffected by vacancies formed relatively far away from the domain wall.

\subsection{Robustness of the chiral states II: Continuous mass inversion}

In practice, mass inversion in gapped graphene is expected to be observed as a consequence of different lattice constants between graphene and other hexagonal substrates. The most common case can be found in graphene on hBN substrates, as follows. In one region, A and B sites of graphene locate onto boron and nitrogen atoms, resulting in a mass gap. In another region, on the other hand, the opposite crystalline configuration between graphene and hBN creates an inverted mass. Of course, one expects the inverted mass to be continuously connected, and the continuous mass inversion depends on how large the lattice mismatch is between graphene and the substrate material.

\begin{figure}
\centering
\includegraphics[width = 8.5cm]{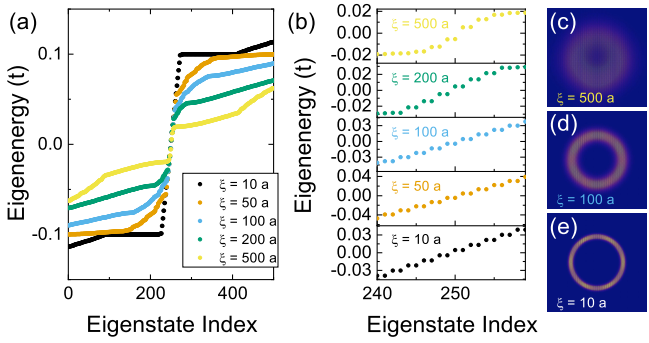}
\caption{Robustness of the chiral states around the MIQD against continuous mass inversion. (a) Eigenenergy spectra for various continuous mass profiles with $\Delta_{0}=0.1~t$. (b) Zoom-in plots of the eigenenergy spectra from (a). (c)--(e) Density maps of selected eigenvectors for $\xi=500$, 100, and 10 $a$, respectively.}
\label{fig:contmass}
\end{figure}

In order to take into account the continuous mass inversion over some distance, the spatial distribution of the mass term is formulated by
\begin{align}
\Delta\left(r\right)=\tanh{\left(\frac{r-r_{0}}{\xi}\right)},
\end{align}
where $\xi$ is the parameter responsible for the steepness of the MIQD boundary. As shown in Fig. \ref{fig:contmass}, the topological edge states around the MIQD are robust against the continuous mass profile. It is also observed in Fig. \ref{fig:contmass}(b) that the chiral symmetry of the topological states is preserved in the continuous MIQD boundary. As displayed in Fig. \ref{fig:contmass}(c)--(e), the broader the chiral channel around the MIQD, the more gradual the mass profile of the MIQD. Meanwhile, as $\xi$ increases, the mass gap narrows since the effective mass gap of the system is related to the averaged mass term:
\begin{align}
E_{g}\approx\left<\sqrt{\Delta\left(r\right)^{2}}\right>.
\end{align}
Such robustness against the mass profile tells us that the chiral states can be induced in graphene on hBN and other two-dimensional (2D)transition metal dichalcogenide (TMDC) substrates, but remain invisible due to the conserved chiral symmetry.

\section{Conclusions}

In this work, a theoretical investigation into a mass-inverted quantum dot in graphene has been carried out to examine the properties of the topological edge states emerging at the mass-inverted domain wall. The numerical results deliver the following implications in terms of both pure and applied physics, informing applications of the topological nature for state-of-art research and technologies.

Recently, myriad studies have focused on the topological states in graphene superlattices or quasicrystals induced by interlayer coupling between substrate crystals, as well as twisted bilayer graphene. The physics of such topological states has not only expanded our understanding of these systems but also revealed the potential for novel nanotechnologies using the topologically protected conducting channels. Through this study, we find that the doubly degenerate chiral states around the MIQD are able to be employed in realizing a stable quantum bit for quantum information technology, based on topology. Topologically protected, a qubit based on an MIQD is advantageous since the chiral states are in principle lossless and barely affected by non-topological defects or disorders.

Lastly, let us mention further research on the robustness of the chiral states around the MIQD. In practice, graphene on a 2D substrate such as hBN or TMDCs experiences periodic modulation of staking order, resulting in an alternating mass-term distribution between carbon and the sublattice atoms of the substrate. Thanks to the recent focus on the bilayers of 2D materials, a continuous distribution of mass gap in graphene would be of interest to researchers in the near future. Further, it is also exciting to explore the effects of periodic topological defects in graphene and other 2D materials.

\section*{Acknowledgments}

The author thanks Joel Rasmussen for the thorough English editing of the manuscript. This work is funded by Chosun University(2019).

\bibliography{MassQDGra}

\begin{thebibliography}{10}
\expandafter\ifx\csname url\endcsname\relax
  \def\url#1{\texttt{#1}}\fi
\expandafter\ifx\csname urlprefix\endcsname\relax\def\urlprefix{URL }\fi
\expandafter\ifx\csname href\endcsname\relax
  \def\href#1#2{#2} \def\path#1{#1}\fi

\bibitem{Kane2005}
C.~L. Kane, E.~J. Mele, Quantum spin hall effect in graphene, Phys. Rev. Lett.
  95 (2005) 226801.

\bibitem{Bernevig2006}
B.~A. Bernevig, T.~L. Hughes, S.-C. Zhang, Quantum spin hall effect and
  topological phase transition in hgte quantum wells, Science 314 (2006)
  1757--1761.

\bibitem{Martin2008}
I.~Martin, Y.~M. Blanter, A.~F. Morpurgo, Topological confinement in bilayer
  graphene, Phys. Rev. Lett. 100 (2008) 036804.

\bibitem{Jung2011}
J.~Jung, F.~Zhang, Z.~Qiao, A.~H. MacDonald, Valley-hall kink and edge states
  in multilayer graphene, Phys. Rev. Lett 84 (2011) 075418.

\bibitem{Yang2012}
F.~Yang, L.~Mioa, Z.~F. Wang, M.-Y. Yao, F.~Zhu, Y.~R. Song, M.-X. Wang, J.-P.
  Xu, A.~V. Fedorov, Z.~Sun, G.~B. Zhang, C.~Liu, F.~Liu, D.~Qian, C.~L. Gau,
  J.-F. Jia, Spatial and energy distribution of topological edge states in
  single bi(111) bilayer, Phys. Rev. Lett. 109 (2012) 016801.

\bibitem{Chang2013}
C.-Z. Chang, J.~Zhang, X.~Feng, J.~Shen, Z.~Zhang, M.~Guo, K.~Li, Y.~Ou,
  P.~Wei, L.-L. Wang, Z.-Q. Ji, Y.~Feng, S.~Ji, X.~Chen, J.~Jia, X.~Dai,
  Z.~Fang, S.-C. Zhang, Y.~Wang, L.~Lu, X.-C. Ma, Q.-K. Xue, Experimental
  observation of the quantum anomalous hall effect in a magnetic topological
  insulator, Science 340 (2013) 167--170.

\bibitem{Vaezi2013}
A.~Vaezi, Y.~Liang, D.~H. Ngai, L.~Yang, E.-A. Kim, Topological edge states at
  a tilt boundary in gated multilayer graphene, Phys. Rev. X 3 (2013) 021018.

\bibitem{Zhang2013}
F.~Zhang, A.~H. MacDonald, E.~J. Mele, Valley chern numbers and boundary modes
  in gapped bilayer graphene, Proc. Nat. Acad. Sci. 110 (2013) 10546--10551.

\bibitem{Drozdov2014}
I.~K. Drozdov, A.~Alexandradinata, S.~Jeon, S.~Nadj-Perge, H.~Ji, R.~J. Cava,
  B.~A. Berevig, A.~Yazdani, One-dimensional topological edge states of bismuth
  bilayers, Nat. Phys. 10 (2014) 664--669.

\bibitem{Li2014}
X.~Li, F.~Zhang, Q.~Niu, A.~H. MacDonald, Spontaneous layer-pseudospin domain
  walls in bilayer graphene, Phys. Rev. Lett. 113 (2014) 116803.

\bibitem{Qiao2014}
Z.~Qiao, J.~Jung, C.~Lin, Y.~Ren, A.~H. MacDonald, Q.~Niu, Current partition at
  topological channel intersections, Phys. Rev. Lett. 112 (2014) 206601.

\bibitem{Du2015}
L.~Du, I.~Knez, G.~Sullivan, R.-R. Du, Robust helical edge transport in gated
  inas/gasb bilayers, Phys. Rev. Lett. 114 (2015) 096802.

\bibitem{Ju2015}
L.~Ju, Z.~Shi, N.~Nair, Y.~Lv, C.~Jin, J.~Velasco~Jr., C.~Ojeda-Aristizabal,
  H.~A. Bechel, M.~C. Martin, A.~Zettl, J.~Analytis, F.~Wang, Topological
  valley transport at bilayer graphene domain walls, Nature(London) 520 (2015)
  650--655.

\bibitem{Li2016}
J.~Li, K.~Wang, K.~J. McFaul, Z.~Zern, Y.~Ren, K.~Watanabe, T.~Taniguchi,
  Z.~Qiao, J.~Zhu, Gate-controlled topological conducting channels in bilayer
  graphene, Nat. Nanotechnol. 11 (2016) 1060--1065.

\bibitem{Zhu2019}
S.-Y. Zhu, Y.~Shao, E.~Wang, L.~Cau, X.-Y. Li, Z.-L. Liu, C.~Liu, L.-W. Liu,
  J.-O. Wang, K.~Ibrahim, J.-T. Sun, Y.-L. Wang, S.~Du, H.-J. Gau, Evidence of
  topological edge states in buckled antimonene monolayers, Nano Lett. 2019
  (2019) 6323--6329.

\bibitem{Hu2019}
X.~Hu, N.~Mao, H.~Wang, Niu, Chengwang, B.~Huang, Y.~Dai, Two-dimensional
  ferroelastic topological insulator with tunable topological edge states in
  single-layer zrasx (x = br and cl), J. Mater. Chem. C 7 (2019) 9743--9747.

\bibitem{Gou2020}
J.~Gou, L.~Kong, X.~He, Y.~L. Huang, J.~Sun, S.~Meng, K.~Wu, L.~Chen, A.~T.~S.
  Wee, The effect of moir\'{e} superstructures on topological edge states in
  twisted bismuthene homojunctions, Sci. Adv. 6 (2020) eaba2773.

\bibitem{Hafezi2013}
M.~Hafezi, S.~Mittal, J.~Fan, A.~Migdall, J.~M. Taylor, Imaging topological
  edge states in silicon photonics, Nat. Photon. 7 (2013) 1001--1005.

\bibitem{Barik2016}
S.~Barik, H.~Miyake, W.~DeGottardi, E.~Waks, M.~Hafezi, Two-dimensionally
  confined topological edge states in photonic crystals, New J. Phys. 18 (2016)
  113013.

\bibitem{StJean2017}
P.~St-Jean, V.~Goblot, E.~Galopin, A.~Lema\^{i}tre, T.~Ozawa, L.~Le~Gratiet,
  I.~Sagnes, J.~Bloch, A.~Amo, Lasing in topological edge states of a
  one-dimensional lattice, Nat. Photon. 11 (2017) 651--656.

\bibitem{Wu2017}
X.~Wu, Y.~Meng, J.~Tian, Y.~Huang, H.~Xian, D.~Han, W.~Wen, Direct observation
  of valley-polarized topological edge states in designer surface plasmon
  crystals, Nat. Commun. 8 (2017) 1304.

\bibitem{Xiao2017}
L.~Xiao, X.~Zhan, Z.~H. Bian, K.~K. Wang, X.~Zhang, X.~P. Wang, J.~Li,
  K.~Mochizuki, D.~Kim, N.~Kawakami, W.~Yi, H.~Obuse, B.~C. Sanders, P.~Xue,
  Observation of topological edge states in parity–time-symmetric quantum
  walks, Nat. Phys. 13 (2017) 1117--1123.

\bibitem{Noh2018}
J.~Noh, S.~Huang, K.~P. Chen, M.~C. Rechtsman, Observation of photonic
  topological valley hall edge states, Phys. Rev. Lett. 120 (2018) 063902.

\bibitem{Han2019}
C.~Han, M.~Lee, S.~Callard, C.~Seassal, H.~Jeon, Lasing at topological edge
  states in a photonic crystal l3 nanocavity dimer array, Light Sci. Appl. 8
  (2019) 40.

\bibitem{Olekhno2020}
N.~A. Olekhno, E.~I. Kretov, A.~A. Stepanenko, P.~A. Ivanova, V.~V. Yaroshenko,
  E.~M. Puhtina, D.~S. Filonov, B.~Cappello, L.~Matekovits, M.~A. Gorlach,
  Topological edge states of interacting photon pairs emulated in a
  topolectrical circuit, Nat. Commun. 11 (2020) 1436.

\bibitem{Yu2020}
Y.~Yu, W.~Song, C.~Chen, T.~Chen, H.~Ye, X.~Shen, Q.~Cheng, T.~Li, Phase
  transition of non-hermitian topological edge states in microwave regime,
  Appl. Phys. Lett. 116 (2020) 211104.

\bibitem{Ni2019}
X.~Ni, K.~Chen, M.~Weiner, D.~J. Apigo, C.~Prodan, A.~Al\'{u}, E.~Prodan, A.~B.
  Khanikaev, Observation of hofstadter butterfly and topological edge states in
  reconfigurable quasi-periodic acoustic crystals, Commun. Phys. 2 (2019) 55.

\bibitem{Gao2020}
H.~Gao, H.~Xue, Q.~Wang, Z.~Gu, T.~Liu, J.~Zhu, B.~Zhang, Observation of
  topological edge states induced solely by non-hermiticity in an acoustic
  crystal, Phys. Rev. B 101 (2020) 180303(R).

\bibitem{Wang2020}
J.~Wang, N.~Gao, H.~Zhang, X.~Zhou, C.~L\"{u}, W.~Chen, Experimentally
  tailoring acoustic topological edge states by selecting the boundary type,
  Appl. Phys. Lett. 117 (2020) 033503.

\bibitem{Xia2020}
Y.~Xia, A.~Erturk, M.~Ruzzene, Topological edge states in quasiperiodic locally
  resonant metastructures, Phsy. Rev. Appl. 13 (2020) 014023.

\bibitem{Goldman2013}
N.~Goldman, J.~Dalibard, A.~Dauphin, F.~Gerbier, M.~Lewenstein, P.~Zoller,
  I.~B. Spielman, Direct imaging of topological edge states in cold-atom
  systems, Proc. Nat. Acd. Sci. 110 (2013) 6736--6741.

\bibitem{Bauer2018}
D.~Bauer, K.~K. Hansen, High-harmonic generation in solids with and without
  topological edge states, Phys. Rev. Lett. 120 (2018) 177401.

\bibitem{Liu2019a}
F.~Liu, H.-Y. Deng, K.~Wakabayashi, Helical topological edge states in a
  quadrupole phase, Phys. Rev. Lett. 122 (2019) 086804.

\bibitem{Obana2019}
D.~Obana, F.~Liu, K.~Wakabayashi, Topological edge states in the
  su-schrieffer-heeger model, Phys. Rev. B 100 (2019) 075437.

\bibitem{Smirnova2019}
D.~A. Smirnova, L.~A. Smirnov, D.~Leykam, Y.~S. Kivshar, Topological edge
  states and gap solitons in the nonlinear dirac model, Laser Photonics Rev. 13
  (2019) 1900223.

\bibitem{Zhang2019}
W.~Zhang, X.~Chen, Y.~V. Kartashov, V.~V. Konotop, F.~Ye, Coupling of edge
  states and topological bragg solitons, Phys. Rev. Lett. 123 (2019) 254103.

\bibitem{Ezawa2020}
M.~Ezawa, Non-abelian braiding of majorana-like edge states and topological
  quantum computations in electric circuits, Phys. Rev. B 102 (2020) 075424.

\bibitem{Song2020}
A.~Y. Song, X.-Q. Sun, A.~Dutt, M.~Minkov, C.~Wojcik, H.~Wang, I.~A.~D.
  Williamson, M.~Orenstein, S.~Fan, Pt-symmetric topological edge-gain effect,
  Phys. Rev. Lett. 125 (2020) 033603.

\bibitem{Yin2016}
L.-J. Yin, H.~Jian, J.-B. Qiao, L.~He, Direct imaging of topological edge
  states at a bilayer graphene domain wall, Nat. Commun. 7 (2016) 11760.

\bibitem{Liu2019b}
J.~Liu, J.~Liu, X.~Dai, Pseudo landau level representation of twisted bilayer
  graphene: Band topology and implications on the correlated insulating phase,
  Phys. Rev. B 99 (2019) 155415.

\bibitem{Wang2019}
Z.-H. Wang, F.~Xu, L.~L\"{u}, B.~Wang, W.-Q. Chen, One-dimensional topological
  superconductivity at the edges of twisted bilayer graphene nanoribbons, Phys.
  Rev. B 100 (2019) 094531.

\bibitem{Ma2020}
C.~Ma, Q.~Wang, S.~Milis, X.~Chen, B.~Deng, S.~Yuan, C.~Li, K.~Watanabe,
  T.~Taniguchi, X.~Du, F.~Zhang, F.~Xia, Moir\'{e} band topology in twisted
  bilayer graphene, Nano Lett. 20 (2020) 6076--6083.

\bibitem{Jiang2020}
L.~Jiang, S.~Wang, S.~Zhao, M.~Crommie, F.~Wang, Soliton-dependent electronic
  transport across bilayer graphene domain wall, Nano Lett. 20 (2020)
  5936--5942.

\bibitem{Chou2020}
Y.-Z. Chou, F.~Wu, S.~Das~Sarma, Hofstadter butterfly and floquet topological
  insulators in minimally twisted bilayer graphene, Phys. Rev. Res. 2 (2020)
  033271.

\bibitem{Lado2013}
J.~L. Lado, J.~W. Gonz\'{a}lez, J.~Fern\'{a}ndez-Rossier, Quantum hall effect
  in gapped graphene heterojunctions, Phys. Rev. B 88 (2013) 035448.

\bibitem{Ratnikov2009}
P.~V. Ratnikov, Superlattice based on graphene on a strip substrate, JETP Lett.
  90 (2009) 469.

\bibitem{Rusponi2011}
S.~Rusponi, M.~Papagno, P.~Moras, S.~Vlaic, M.~Etzkorn, P.~M. Sheverdyaeva,
  D.~Pacil\'{e}, H.~Brune, C.~Carbone, Highly anisotropic dirac cones in
  epitaxial graphene modulated by an island superlattice, Phys. Rev. Lett. 105
  (2011) 246803.

\bibitem{Kim2018}
H.~Kim, N.~Leconte, B.~L. Chittari, K.~Watanabe, T.~Taniguchi, A.~H. MacDonald,
  J.~Jung, S.~Jung, Accurate gap determination in monolayer and bilayer
  graphene/h-bn moir\'{e} superlattices, Nano Lett. 18 (2018) 7732–--7741.

\bibitem{VanPottelberge2017}
R.~Van~Pottelberge, M.~Zarenia, P.~Vasilopoulos, F.~M. Peeters, Graphene
  quantum dot with a coulomb impurity: Subcritical and supercritical regime,
  Phys. Rev. B 95 (2017) 245410.

\bibitem{Belouad2018}
A.~Belouad, Y.~Zahidi, A.~Jellal, H.~Bahlouli, Electron scattering in gapped
  graphene quantum dots, Eur. Phys. Lett. 123 (2018) 28002.

\bibitem{Xu2020}
H.-Y. Xu, Y.-C. Lai, Anomalous chiral edge states in spin-1 dirac quantum dots,
  Phys. Rev. Res. 2 (2020) 013062.

\bibitem{Groth2014}
C.~W. Groth, M.~Wimmer, A.~R. Akhmerov, X.~Waintal, Kwant: a software package
  for quantum transport, New. J. Phys. 16 (2014) 063065.

\bibitem{Wang2014}
J.~Wang, S.~Fischer, Topological valley resonance effect in graphene, Phys.
  Rev. B 89 (2014) 245421.

\bibitem{Weisse2008}
A.~Wei{\ss}e, G.~Wellein, A.~Alvermann, H.~Fehske, The kernel polynomial
  method, Rev. Mod. Phys. 78 (2006) 275.

\bibitem{Libisch2011}
F.~Libisch, S.~Rotter, J.~Burgd\"{o}rfer, Disorder scattering in
  graphenenanoribbons, Phys. Status Solidi B 248 (2011) 2598–2603.

\bibitem{RuizTijerina2016}
D.~A. Ruiz-Tijerina, L.~G. G.~V. Dias~da Silva, Symmetry-protected coherent
  transport for diluted vacancies and adatoms in graphene, Phys. Rev. B 94
  (2016) 085425.

\end{thebibliography}

\end{document}